\RequirePackage{fix-cm}
\documentclass[smallextended]{svjour3}       
\smartqed  
\usepackage{graphicx}
\usepackage{amsmath}
 \usepackage{mathptmx}      
\usepackage{latexsym}

\usepackage{physics}

\newcommand{\Beq}{\begin{equation}}
\newcommand{\Eeq}{\end{equation}}
\renewcommand{\vec}[1]{\boldmath{#1}}

\begin{document}

\title{Relativistic Feshbach-Villars Equation for Two Spin-$0$ Particles 
}


\author{  Z.~Papp
}


\institute{              
                Z.~Papp
	\at
              Department of Physics  and Astronomy \\
              California State University Long Beach \\
              California, USA\\
              \email{Zoltan.Papp@csulb.edu} 	      
}

\date{Received: date / Accepted: date}

\maketitle

\begin{abstract}
The Feshbach-Villars version of the relativistic quantum mechanics  
 can be extended for two-body systems in such a way that the 
 center-of-mass motion is separated off. The procedure results in an equation of
 Feshbach-Villars-type in terms of the relative coordinate.

\end{abstract}

\section{Introduction }

The basic equations of the relativistic quantum mechanics, the Klein-Gordon (KG0) 
equation for spin-$0$ and the Dirac equation for spin-$1/2$ particles, 
are  one-particle equations or equations for
quantum fields. 

Moreover, the Klein-Gordon equation also contradicts the basic postulates of quantum mechanics. 
In quantum mechanics, it is postulated that the system is 
completely determined by the wave function and the time evolution of the wave function is governed
by the time-dependent  Schr\"odinger equation. The Klein-Gordon
equation is second order in time derivative. 
Therefore, to determine the system uniquely,  we need its time derivative as well. 
Also, the stationary form of the Klein-Gordon equation contains the square of the energy.
This prevents us developing a consistent few-particle theory since we cannot add the energies of independent 
particles to get the energy of the whole system.  So, the Klein-Gordon equation is strictly a one-particle equation.

In order to give a proper interpretation to the Klein-Gordon equation, Feshbach and Villars rewrote it in a Hamiltonian form
 \cite{Feshbach:1958wv}. In  the Feshbach-Villars (FV0)  formalism we split
the Klein-Gordon wave function into two components, and  for the components we
obtain a Schr\"odinger-like equation with a first order time derivative.  
The formalism offers an interesting interpretation of relativistic  particles. A relativistic particle is a mixture of
particle and antiparticle components, and the components are coupled by the kinetic energy. 
There are no pure particle or antiparticle states,
only one component is dominant over the other.
The quantum system is more like a particle or more like an antiparticle. It became pure particle or antiparticle when it is at rest.

The kinetic energy coupling makes the solution of the Feshbach-Villars equation notoriously difficult,
as it couples the components even at asymptotic distances. The components do not get decoupled even 
asymptotically.
Nevertheless, in a recent work we have proposed a solution method for the Feshbach-Villars equation
\cite{brown2016matrix,motamedi2019relativistic}. 
The method casts the eigenvalue problem 
into a Lippmann-Schwinger  equation  and represents the interaction part of the Hamiltonian on a discrete Hilbert-space basis. 
The corresponding Feshbach-Villars Green's operator has been calculated by a matrix continued fraction.

This paper is an attempt to generalize the Feshbach-Villars equation for two-body systems. 
We show that this formalism allows the separation of the center-of-mass kinetic energy, which is 
a crucial point in establishing a consistent few-body theory. 
In section 2 we outline the Feshbach-Villars equation, in Section 3 we extend it to two-body systems, in Section 4
we calculate the low-lying states of $p-\pi^{-}$ and the $\pi^{+}-\pi^{-}$ Coulomb systems, and, finally, 
in Section 5 we summarize our findings.

\section{Feshbach-Villars  equation for spin-zero particles}
\label{fv0-outline}

The Klein-Gordon equation for a free spin-$0$ particle with mass $m$ is given by
\begin{equation}
-\hbar^{2} \frac{\partial^{2}}{\partial t^{2}}    \Psi  = \left( c^{2} p^{2} + m^{2}c^{4} \right)  \Psi .
\end{equation}
We can introduce interaction by minimal coupling,  $p_{\mu} \to p_{\mu}  -  q/c\,  A_{\mu}$, where $p_{\mu}$ and $A_{\mu}$ are the 
four-momentum and the four-potential, respectively. This interaction transforms like a four-vector with respect to the
Lorentz transformation. 
We can also introduce a scalar interaction $S$ by the substitution $m \to m + S/c^{2}$, 
which is basically a position dependent effective mass.
 So, if we take $\vec{A}=0$ and denote the time-like component of the vector potential by $V$, we have
\begin{equation}
\left(i\hbar  {\partial}/{\partial t}  - V \right)^{2}  \Psi  = \left[ c^{2} p^{2} + (m + S/c^{2})^{2}c^{4} \right]  \Psi .
\end{equation}
Here, $V$ is the time-like component of a Lorentz four-vector, like the Coulomb or the Yukawa potential. 
The scalar potential $S$ effectively modifies the rest mass of the particle. It can be used to mimic the interaction with the Higgs field, 
the effect of the surrounding media, or the quark confinement.

In the Feshbach-Villars formalism the wave function is split into two components
\begin{equation}
\Psi =  \phi + \chi \quad \text{and} \quad
\left( i \hbar \frac{\partial }{ \partial t} - V \right) \Psi  =  m c^{2} (\phi -\chi),
\end{equation}
such that
\begin{equation}
\phi =   \frac{1}{2} \left[ 1+ \frac{1}{m c^{2}} \left( i \hbar  \frac{\partial}{\partial t } - V \right) \right] \Psi \quad \text{and} \quad
\chi   =   \frac{1}{2} \left[ 1- \frac{1}{m c^{2}} \left( i \hbar  \frac{\partial}{\partial t } - V \right) \right] \Psi. \label{phichi}
\end{equation}

For the components we can readily obtain the coupled equations
\begin{eqnarray}
i\hbar \frac{\partial}{\partial t} \phi & = & \left(  \frac{p^{2}}{2m} +U \right)  (\phi + \chi) + ( mc^{2} +V) \phi ~,  \nonumber \\
i\hbar \frac{\partial}{\partial t} \chi & = & - \left(  \frac{p^{2}}{2m} +  U \right) (\phi + \chi) - ( mc^{2} -V) \chi   ~,  \label{fv2}  
\end{eqnarray}
where $U = S + {S^{2}}/({2mc^{2}})$.
These equations for stationary problems take the form
\begin{eqnarray}
 \left(  \frac{p^{2}}{2m} +U \right)  (\phi + \chi) + ( mc^{2} +V) \phi  & = & E \phi ~,  \nonumber \\
 - \left(  \frac{p^{2}}{2m} +  U \right) (\phi + \chi) - ( mc^{2} -V) \chi  & = &E \chi ~.  \label{fv22}  
\end{eqnarray}

If we introduce the two-component wave function
\begin{equation}
| \psi \rangle = \begin{pmatrix}  \phi \\ \chi \end{pmatrix} ,
\end{equation}
we can define the Hamiltonian
\begin{equation} \label{HFV0}
H_{FV0}  =  (\tau_{3}+i\tau_{2}) \frac{p^{2}}{2m} +\tau_{3} mc^{2} + (\tau_{3}+i\tau_{2}) U + I_{2} V ,
\end{equation}
where $\tau_{i}$  denote the Pauli matrices 
\begin{equation}
\tau_{1}= \begin{pmatrix}  0 & 1 \\ 1 & 0 \end{pmatrix}, \ \ \ \tau_{2}= \begin{pmatrix}  0 & -i \\ i & 0 \end{pmatrix}, \ \ \
\tau_{3}= \begin{pmatrix}  1 & 0 \\ 0 & -1 \end{pmatrix},
\end{equation}
that act on the Feshbach-Villars components and $I_{2}$ is the $2 \times 2$ identity matrix.
Now we can write  Eqs.\ (\ref{fv2}) into a form analogous to the time-dependent Schr\"odinger equation
\begin{equation}
i \hbar \frac{\partial}{\partial t} |\psi \rangle = H_{FV0} | \psi \rangle,
\end{equation}
or, for stationary states, we have
\begin{equation}
  H_{FV0} | \psi \rangle = E | \psi \rangle ~.
  \label{hpsi}
\end{equation}

We can gain some insight into the meaning of the components by taking Eqs.\ (\ref{phichi}) with $i\hbar \partial/ \partial t \rightarrow E$ and $V=0$
\begin{equation}
\phi =   \frac{1}{2} \left[ 1+ \frac{E}{m c^{2}}   \right] \Psi \quad \text{and} \quad
\chi   =   \frac{1}{2} \left[ 1- \frac{E}{m c^{2}}   \right] \Psi. \label{phichi1}
\end{equation}
We can see that for positive energies the component $\phi$ dominate over the component $\chi$, while for negative energies the situation is opposite.
In general, a quantum state in the Feshbach-Villars formalism is a combination of $\phi$ and $\chi$ components, where $\phi$ measures
the particle content and $\chi$ measures the antiparticle content. These components are separated by the energy $2 mc^{2}$ 
and coupled by the kinetic energy $p^{2}/(2m)$. In the low energy limit, where $E \simeq mc^{2}$,  
the component $\phi$ is basically the total Klein-Gordon wave function $\Psi$,
$\phi \simeq \Psi$, and the component $\chi$ is diminishing,  $\chi \simeq 0$. 
At high energies as $E \to \infty$ we find that the two components have more or less
the same weight, $\phi \sim \chi$. The equation has solution at negative energies as well, where the role of $\phi$ and $\chi$ is interchanged.

The Hamiltonian $H_{FV0}$ of Eq.\ (\ref{HFV0}) is not
Hermitian in the usual sense, it is Hermitian in the generalized sense 
\begin{equation}
H_{FV0} = \tau_{3} H_{FV0}^{\dagger} \tau_{3},
\end{equation}
and it possesses real eigenvalues  \cite{Feshbach:1958wv,wachter2010relativistic}. The wave function is normalized according to
\begin{equation}
\langle \psi | \tau_{3} | \psi \rangle = \pm 1,
\end{equation}
where the plus or minus sign corresponds to particle or antiparticle.

Eqs.\ (\ref{fv2}) look like a usual set of coupled channel equations. 
However, usually 
the channels are coupled by some short-range potential that vanishes at asymptotic distances. 
Here the coupling is due to the kinetic energy operator, which is not
a short-range operator and cannot be neglected even at asymptotic distances. 
Moreover, the determinant of $\tau_{3}+i\tau_{2}$ vanishes,
so the coupling cannot be removed by multiplying with the inverse.

 \section{Feshbach-Villars equation for two particle systems}

Let us consider two spin-$0$ particles $\alpha$ and $\beta$, with masses $m_{\alpha}$ and $m_{\beta}$, 
located at $\vec{r}_{\alpha}$ and $\vec{r}_{\beta}$, respectively. 
We can introduce the relative and the center-of-mass coordinates 
\begin{equation}
\vec{r} = \vec{r}_{\alpha} - \vec{r}_{\beta} \quad \text{and} \quad 
\vec{R} = \frac{m_{\alpha}  r_{\alpha}+m_{\beta} r_{\beta}}{m_{\alpha} +m_{\beta}}.
\end{equation}
The total and the reduces masses are given by
\begin{equation}
M = m_{\alpha}+m_{\beta} \quad \text{and} \quad \mu = m_{\alpha} m_{\beta}/M,
\end{equation}
respectively. With $r_{\alpha} = R  + m_{\beta} r /M $ and $r_{\beta} = R -m_{\alpha} r /M $ we easily find that
\begin{equation}
\frac{1}{2}m_{\alpha} \dot{r}_{\alpha}^{2} + \frac{1}{2}m_{\beta} \dot{r}_{\beta}^{2}  = \frac{1}{2} M \dot{R}^{2} + \frac{1}{2} \mu \dot{r}^{2},
\end{equation}
i.e.\  the non-relativistic kinetic energy of a two-particle system can be written as the sum of the center-of-mass kinetic energy and
the kinetic energy in the relative coordinate with reduced mass $\mu$.

The Feshbach-Villars equation is a genuine eigenvalue equation for the energy.
The energies of the two free spin-$0$ particles are determined  by 
\begin{equation}
E_{\alpha} \ket{\Psi_{\alpha} }  = 
\left(  (\tau_{3}+i\tau_{2}) \frac{p_{\alpha}^{2}}{2m_{\alpha}} +\tau_{3} m_{\alpha}c^{2} \right)  \ket{\Psi_{\alpha} } 
\end{equation}
and 
\begin{equation}
E_{\beta} \ket{\Psi_{\beta} }  = \left(  (\tau_{3}+i\tau_{2}) \frac{p_{\beta}^{2}}{2m_{\beta}} +\tau_{3} m_{\beta}c^{2} \right)  \ket{\Psi_{\beta} },
\end{equation}
respectively. 
The energy is additive, therefore the equation for the unified system reads
\Beq
E \ket{ \Psi } = \left[  (\tau_{3}+i\tau_{2}) \left( \frac{p_{\alpha}^{2}}{2m_{\alpha}}  + \frac{p_{\beta}^{2}}{2m_{\beta}} \right)   
+  \tau_{3}  ( m_{\alpha}c^{2} + m_{\beta}c^{2} ) \right] \ket{\Psi},
\Eeq
with $E = E_{\alpha} + E_{\beta}$ and $\Psi (r_{\alpha}, r_{\beta})= \Psi_{\alpha}(r_{\alpha})  \Psi_{\beta}(r_{\beta})$. 

Introducing an interaction into a free Hamiltonian is a subtle issue. 
However, in quantum mechanics we assume that the interaction on the particle $\alpha$ 
can come only from particle $\beta$, and vice versa. So, it is plausible to assume 
 that the mutual interaction of the particles depends only on their
relative coordinates. Consequently,  the Hamiltonian becomes
\Beq
H =  (\tau_{3}+i\tau_{2}) \left( \frac{p_{\alpha}^{2}}{2m_{\alpha}}  + \frac{p_{\beta}^{2}}{2m_{\beta}} \right)   
+  \tau_{3}  ( m_{\alpha}c^{2} + m_{\beta}c^{2} )  +  (\tau_{3}+i\tau_{2}) U(\vec{r}) + I_{2}V(\vec{r}).
\label{hami0}
\Eeq

As we know from the non-relativistic quantum mechanics, the kinetic energy of a two-particle system 
can be expressed in terms of center of mass coordinates
\Beq
\frac{p_{\alpha}^{2}}{2m_{\alpha}}  + \frac{p_{\beta}^{2}}{2m_{\beta}} = \frac{P^{2}}{2M} + \frac{p^{2}}{2\mu},
\Eeq
where  $P=-i\hbar \pdv*{R}$ is the total momentum 
and $p=-i\hbar \pdv*{r}$ is the relative momentum. So, Eq.\ (\ref{hami0}) becomes
\Beq
H =   (\tau_{3}+i\tau_{2})  \frac{P^{2}}{2M}    +        (\tau_{3}+i\tau_{2})   \frac{p^{2}}{2\mu} +  \tau_{3}   M c^{2} 
 +  (\tau_{3}+i\tau_{2}) U(\vec{r}) + I_{2}V(\vec{r}),
\label{hami01}
\Eeq
and the eigenvalue equation takes the form
\begin{equation}
H\ket{\psi} = E \ket{\psi}
\end{equation}
with $\ket{\psi(R,r)} = \ket{\psi_{R}(R)} \ket{\psi_{r}(r)}$,  a product Feshbach-Villars wave functions in coordinates $R$ and $r$, respectively. 
We can see that the first term represents  the kinetic energy of the center-of-mass motion.
This term does not carry much physics and it can be eliminated by putting the reference frame to the center-of-mass of the system. 
What remains is the Hamiltonian for the relative motion
\Beq
H_{r} = (\tau_{3}+i\tau_{2})   \frac{p^{2}}{2\mu} + \tau_{3}   M c^{2}
 +  (\tau_{3}+i\tau_{2}) U(\vec{r}) + I_{2} V(\vec{r}).
 \label{fv0rel}
\Eeq
This is just like the  Feshbach-Villars Hamiltonian Eq.\ (\ref{HFV0}) in terms of the relative coordinate $r$,
with the effective mass $\mu$ in the kinetic energy term and the total mass $M$ in the rest energy term.
The $\tau_{3} M c^{2}$  sets the separation of the energy levels between particle and antiparticle states,  
in a nice agreement with the non-relativistic limit.

\section{Solution method} 

The main difficulty in solving the Hamiltonian in Eq.\ (\ref{fv0rel}) is the handling of the coupling by the kinetic energy. 
This coupling cannot be removed.
We cannot multiply the equation by the inverse of $\tau_{3}+i \tau_{2}$, since the inverse does not exist as the determinant vanishes.  
A differential equation solution does not seem to be a viable approach.
Therefore, in Ref.\ \cite{motamedi2019relativistic} we worked out a solution method for Hamiltonians like in Eq.\ (\ref{fv0rel})
for bound and resonant states with short-range plus Coulomb or confining potentials. 

In this approach we split the Hamiltonian into asymptotically relevant long-range and asymptotically irrelevant short-range parts
\begin{equation}
H_{r} = H_{r}^{(l)}+ H_{r}^{(s)}.
\end{equation}
Here 
\begin{equation}
H_{r}^{(s)} = (\tau_{3} + i \tau_{2}) U^{(s)} + I_{2}V^{(s)}
\end{equation}
contains the short-range parts of the interactions and
\begin{equation}
H_{r}^{(l)} = (\tau_{3} + i \tau_{2}) \frac{p^{2}}{2\mu} + \tau_{3} M c^{2} + (\tau_{3} + i \tau_{2}) U^{(l)} + I_{2} V^{(l)}
\label{Hrl}
\end{equation}
is the long-range part of the Hamiltonian. Then, the Lippmann-Schwinger equation for a bound-state problem reads
\begin{equation}
\ket{\psi_{r}} = G_{r}^{(l)} (E) H_{r}^{(s)} \ket{\psi_{r}},
\label{LS}
\end{equation}
where 
\begin{equation}
G_{r}^{(l)} (E) = (E - H_{r}^{(l)})^{-1}
\end{equation}
is the Green's, or resolvent, operator associated with the long-range Feshbach-Villars Hamiltonian $H_{r}^{(l)}$.

For the approximation method, we adopted the Coulomb-Sturmian basis. For a partial wave $l$, the Coulomb-Sturmian functions
 are given by
\begin{equation}
\braket{r}{nl} = \left( \frac{\Gamma(n+1)}{(\Gamma(n+2l+2)} \right)^{1/2} \exp(-br) (2br)^{l+1} L_{n}^{(2l+1)}(2br),
\end{equation}
where $b$ is a parameter, $n$ is the radial quantum number and $L$ is the associated Laguerre polynomial. 
The Coulomb-Sturmian functions, together with the bi-orthonormal partner $\braket{r}{\tilde{nl}}=  \braket{r}{nl}/r$, 
form a basis, i.e.\ 
\begin{equation}
\braket*{nl}{\tilde{n'l}} = \braket*{\tilde{nl}}{{n'l}} =\delta_{nn'} \quad \text{and} \quad \sum_{n=0}^{\infty} \ket*{nl}\bra*{\tilde{nl}} = 
\sum_{n=0}^{\infty} \ket*{\tilde{nl}}\bra*{{nl}} = I_{l},
\end{equation}
where $I_{l}$ is the unit operator in the angular momentum $l$ subspace.

We approximate the short range operator 
\begin{equation}
H_{r}^{(s)} \approx \sum_{n n'}^{N} \ket*{\tilde{nl}} \mel*{nl}{H_{r}^{(s)} }{n'l} \bra*{\tilde{n'l}} 
\end{equation}
and plug in Eq.\ (\ref{LS}) to get
\begin{equation}
\ket{\psi_{r}} =  \sum_{n n'}^{N} G_{r}^{(l)} (E)  \ket*{\tilde{nl}} \mel*{nl}{H_{r}^{(s)} }{n'l}   \braket*{\tilde{n'l}}{\psi_{r}}.
\label{LS1}
\end{equation}
We should notice that the summation over $n'$ runs only op to $N$. So, in order that we can determine $\braket*{\tilde{n'l}}{\psi_{r}}$
we need to act on Eq.\ (\ref{LS1}) by $\bra*{\tilde{ml}}$ with $m$'s up to $N$ only. As a consequence, Eq.\ (\ref{LS1}) becomes a matrix equation
\begin{equation}
\underline{\psi}_{r} = \underline{G}_{r}^{(l)}(E) \underline{H}_{r}^{(s)} \underline{\psi}_{r},
\end{equation}
where $\underline{\psi}_{r} =  \braket*{\tilde{nl}}{\psi_{r}}$,  $\underline{G}_{r}^{(l)} = \mel*{\tilde{nl}}{G_{r}^{(l)}}{\tilde{n' l}}$ 
and  $\underline{H}_{r}^{(s)}  = \mel*{nl}{H_{r}^{(s)} }{n'l}$. By rearranging, we have
\begin{equation}
\left(  (\underline{G}_{r}^{(l)})^{-1} (E)   -  \underline{H}_{r}^{(s)} \right)\underline{\psi}_{r}  = 0, 
\label{LS2}
\end{equation}
which is a homogeneous algebraic equation. The homogeneous equation is solvable if the determinant vanishes
\begin{equation}
 | (\underline{G}_{r}^{(l)})^{-1} (E)   -  \underline{H}_{r}^{(s)} | = 0,
\end{equation} 
which gives us the energy eigenvalues and the solution of Eq.\ (\ref{LS2}) provides us with the eigenstates.

The calculation of $\underline{H}_{r}^{(s)}$ should be a straightforward, at least numerically, for any reasonable potential
$U^{(s)}$ and $V^{(s)}$. 
The Coulomb-Sturmian basis allows an exact and analytical calculation of matrix elements 
\begin{equation}
\mel{nl}{1/r}{n' l}=\delta_{n n'},
\end{equation}
\begin{equation}
\braket{nl}{n' l}= \begin{cases}  (n+l+1)/b & \text{if $ n = n'$ } \\
		-\sqrt{n'(n'+2l+1)} /(2b) & \text{if $n' = n+1$ }  \\
		-\sqrt{n(n+2l+1)} /(2b) & \text{if $n= n'+1$ }  \\
		0 & \text{otherwise},
		 \end{cases}
\end{equation}
and
\begin{equation}
\mel{nl}{p^{2}}{n' l}= \begin{cases}  (n+l+1) b & \text{if $ n = n'$ } \\
		\sqrt{n'(n'+2l+1)} b/2 & \text{if $n' = n+1$ }  \\
		\sqrt{n(n+2l+1)} b/2 & \text{if $n= n'+1$ }  \\
		0 & \text{otherwise}.
		 \end{cases}
\end{equation}
As a consequence, the operator  $J = \lambda - p^{2}/(2m) - Z/r$, where $\lambda$ is a parameter and $Z$ is the charge number,
 has an infinite symmetric tridiagonal representation. The Green's, or the resolvent, operator is defined by the relation
 \begin{equation}
 J(\lambda) G(\lambda) =    G(\lambda)  J(\lambda) = 1.
 \end{equation}
 It has been shown in Ref.\ \cite{Konya:1997JMP} that a finite $N \times N$ representation of the Green's operator
 associated with infinite symmetric tridiagonal Hamiltonians 
  can be given in the form 
 \begin{equation}
 \underline{G} = ( \underline{J} - \delta_{i N} \delta_{j N} J_{N,N+1} C_{N+1} J_{N+1,N} )^{-1},
 \label{jinv}
 \end{equation}
where $C$ is a continued fraction defined by the relation
\begin{equation}
C_{N+1}=(J_{N+1,N+1} - J_{N+1,N+2}  C_{N+2}  J_{N+2,N+1})^{-1}. 
\label{cfrac}
\end{equation}
So, the inverse of the $N \times N$ Green's matrix is basically the $N \times N$ $J$ matrix 
plus a continued fraction added to the right bottom corner.

In Eq.\ (\ref{Hrl}), in  $H_{r}^{(l)}$, the kinetic energy and the constant terms are tridiagonal in the Coulomb-Sturmian basis 
representation, while if the long range potentials are Coulomb potentials, they are diagonal. So, the operator $J = \lambda - H_{r}^{(l)}$,
in Coulomb-Sturmian basis, is tridiagonal due to the underlying structure of the kinetic energy, the mass term  and the Coulomb potential, 
but each element
of that infinite tridiagonal matrix is, in fact, a $2 \times 2$ matrix due to the Feshbach-Villars structure of the Hamiltonian.
So, $J_{i,j}$ elements in Eqs.\ (\ref{jinv}) and  (\ref{cfrac}) are matrices and the continued fraction becomes a matrix continued fraction. 

We should notice that in this solution method we approximate only the short-range potential $H_{r}^{(s)}$. The fact that we need only
the $N \times N$ matrix elements of the Green's operator comes as the consequence of the applied approximation method. 
The matrix continued fraction 
converges fast for bound state problems, but it can also be continued analytically for scattering state energies  \cite{motamedi2019relativistic}.

\section{The pionic hydrogen and the pionium} 

Here, we consider first the pionic hydrogen, assuming zero spin for the proton, 
with an attractive Coulomb potential $V$. The Coulomb potential is the time-like component of the electromagnetic four-potential $A_{\mu}$. 
We take the fine structure constant $\alpha = e^{2}/(\hbar c)= 0.0072973525643 \simeq 1/137.035999177$. 
We use atomic units such that  $\hbar=1$, $e^{2}=1$, and the mass of the electron $m_{e}=1$. Then the mass of the proton
$m_{p}= 1836.15267 m_{e}$, the mass of the pion $m_{\pi}= 273.13244 m_{e}$.  

Tables \ref{table1} and \ref{table2} show a few low lying states in the $p-\pi^{-}$ and in the $\pi^{+}-\pi^{-}$ systems, respectively.
We can see that the KG0 and FV0 results are different. This is due to the different mass scales. 
If we had the effective KG0 in the form of Eq.\ (\ref{HFV0}), the components were separated by $\tau_{3}\mu c^2$, while in 
Eq.\  (\ref{fv0rel}) they are separated by $\tau_{3}M c^2$. The $\tau_{3}\mu c^{2}$ separation do not have any physical basis
as it violates the non-relativistic limit. The rest mass energy of two particles should be $Mc^{2}$, not $\mu c^{2}$.
 Nevertheless,  we show the $KG0$ results for comparison. The numerical method for evaluating
the matrix continued fraction is robust and allows us to pinpoint small relativistic effects.

\begin{table}[h]
\centering
\caption{Binding energies  of the $p-\pi^{-}$ system in atomic units using the Klein-Gordon (KG0) equation with the reduced 
mass $\mu$ and the Feshbach-Villars (FV0) 
equation (Eq.\ (\ref{fv0rel})) .}
\label{table1}
\begin{tabular}{| c | c | c | c | c | c | c | } 
\hline
 & \multicolumn{3}{|c|}{KG0} & \multicolumn{3}{|c|}{FV0} \\
\hline
 $n$ & $l=0$ & $l=1$ & $l=2$  & $l=0$ & $l=1$ & $l=2$       \\ 
 \hline
  1 & -118.890102 &  &    & -118.883080 &   &   \\
  2 &   -29.7218331 & -29.7207779 &   & -29.7206924 & -29.7205735  &   \\
   3 &  -13.2095424 &  -13.2092297  & -13.2091672  & -13.2091785  & -13.2091433   & -13.2091362   \\
\hline
\end{tabular} 
\end{table}

\begin{table}[h]
\centering
\caption{Binding energies  of the $\pi^{+}-\pi^{-}$ system in atomic units using the Klein-Gordon (KG0) equation with the reduced mass $\mu$ 
and the Feshbach-Villars (FV0)
equation (Eq.\ (\ref{fv0rel})) .}
\label{table2}
\begin{tabular}{| c | c | c | c | c | c | c | } 
\hline
 & \multicolumn{3}{|c|}{KG0} & \multicolumn{3}{|c|}{FV0} \\
\hline
 $n$ & $l=0$ & $l=1$ & $l=2$  & $l=0$ & $l=1$ & $l=2$       \\ 
 \hline
  1 & -68.2876557 &  &    & -68.2842463   &   &   \\
  2 &   -17.0715161 & -17.0709101 &   & -17.0709623  & -17.0708110   &   \\
   3 &  -7.5872479  &  -7.5870683  & -7.5870324 & -7.5870713  &  -7.5870264     &   -7.5870174   \\
\hline
\end{tabular} 
\end{table}

\section{Summary and conclusions}

The Feshbach-Villars Hamiltonian Eq.\ (\ref{HFV0}) looks very much like a non-relativistic Hamiltonian,
and unlike the Klein-Gordon equation, it is a genuine eigenvalue equation. 
Another  difference is that it is a matrix equation and the 
components are coupled by the  kinetic energy  via the $2 \times 2$ matrix $\tau_3+i \tau_2$. 
If the energy is positive and kinetic energy is small, $\phi$, which represent the particle component, 
is dominant over $\chi$, which represents the antiparticle component, and the formalism falls back to the non-relativistic 
Schr\"odinger equation.

The Feshbach-Villars equation is an eigenvalue equation for the energy.
As a consequence,  we can express the total energy of non-interacting spin-$0$ particles as
the sum of subsystem energies. The similarity to the non-relativistic kinetic energy naturally allows the separation of the center-of-mass motion,
which remains true even if we add interactions that depend on the relative coordinate $r$. 
As a result, we arrive at an equation in terms of the 
relative coordinate which is very similar to the original Feshbach-Villars equation.  We believe that this result can be
extended to particles with spin, and opens a pathway to establish a consistent relativistic theory for few-particle systems.

In establishing relativistic few-particle equations one may face three major challenges. One issue is the retardation. 
Similarly to the Li\'enard-Wiechert potential in classical electrodynamics, the potential must account for the fact that 
for the exchange boson it takes time to travel between particles. 
However, this problem may not be relevant if we consider stationary eigenvalue 
problems.

Another main issue is the handling of antiparticle solutions. In the Feshbach-Villars approach this is not a problem. The
particle and antiparticle components  show up  explicitly in the formalism. 

Probably the most crucial problem is to ensure the cluster separability, that isolated subsystems behave independently.
In this regard it is paramount that the physics of the subsystem should 
not depend on the relative motion of the outside observer, i.e.\ we should be able to separate off the center-of-mass motion. 
Some approaches achieve this goal by imposing extra constrains. In this paper we show that in the Feshbach-Villars formalism 
the separation of the center-of-mass motion comes naturally and the resulting equations are effective 
Feshbach-Villars equations in the relative coordinate.


\bibliographystyle{spphys}       
\bibliography{fv00}

%
%

\end{document}